# SARS-CoV-2's closest relative, RaTG13, was generated from a bat transcriptome not a fecal swab: implications for the origin of COVID-19

Steven E Massey

stevenEmassey@gmail.com

Department of Biology, University of Puerto Rico - Rio Piedras,
San Juan, Puerto Rico 00901

**Abstract**

RaTG13 is the closest related coronavirus genome phylogenetically to SARS-CoV-2, consequently understanding its provenance is of key importance to understanding the origin of the COVID-19 pandemic. The RaTG13 NGS dataset is attributed to a fecal swab from the intermediate horseshoe bat *Rhinolophus affinis*. However, sequence analysis reveals that this is unlikely. Metagenomic analysis using Metaxa2 shows that only 10.3 % of small subunit (SSU) rRNA sequences in the dataset are bacterial, inconsistent with a fecal sample, which are typically dominated by bacterial sequences. In addition, the bacterial taxa present in the sample are inconsistent with fecal material. Assembly of mitochondrial SSU rRNA sequences in the dataset produces a contig 98.7 % identical to *R.affinis* mitochondrial SSU rRNA, indicating that the sample was generated from this or a closely related species. 87.5 % of the NGS reads map to the *Rhinolophus ferrumequinum* genome, the closest bat genome to *R.affinis* available. In the annotated genome assembly, 62.2 % of mapped reads map to protein coding genes. These results clearly demonstrate that the dataset represents a *Rhinolophus* sp. transcriptome, and not a fecal swab sample. Overall, the data show that the RaTG13 dataset was generated by the Wuhan Institute of Virology (WIV) from a transcriptome derived from *Rhinolophus* sp. tissue or cell line, indicating that RaTG13 was in live culture. This raises the question of whether the WIV was culturing additional unreported coronaviruses closely related to SARS-CoV-2 prior to the pandemic. The implications for the origin of the COVID-19 pandemic are discussed.

## Introduction

Understanding the origin of severe acute respiratory syndrome coronavirus 2 (SARS-CoV-2) and coronavirus disease 2019 (COVID-19) is vital for preventing future pandemics. There are two main hypotheses regarding the origin of the COVID-19 pandemic. The zoonosis hypothesis proposes that the progenitor of SARS-CoV-2 jumped from a bat or intermediate host to a human (1). This scenario requires that the infected bat or intermediate host came into close contact with a human in a non-research setting which allowed the transmission to occur. The contrasting lab leak hypothesis proposes that SARS-CoV-2 was transmitted into the human population from a research related activity such as a laboratory experiment (2).

The RaTG13 coronavirus genome, sequenced by the Wuhan Institute of Virology (WIV), is phylogenetically the closest known relative to SARS-CoV-2[1], and its apparent provenance from the intermediate horseshoe bat *Rhinolophus affinis* has been used to support the proposed zoonotic origin of SARS-CoV-2 (3). However, the original Nature publication describing the RaTG13 genome sequence was sparse regarding sampling location and date of sequencing of RaTG13. A fragment of the RNA dependent RNA polymerase (RdRp) was the first part of RaTG13 to be sequenced, initially labelled as 'RaBtCov/4991' (4), and subsequently renamed 'RaTG13' in the Nature paper (the link between the two was identified in (5)). Further details were elaborated in an Addendum (6), which gave the date of sequencing of RaTG13 as 2018, and the sampling location as a mine in Mojiang, Yunnan Province, China, which had been associated with the death in 2012 of three miners who had been clearing bat guano, from a virus like respiratory infection (7) (8) (9).

[1] While RaTG13 shows 96.2 % sequence identity with the SARS-CoV-2 genome (3), a new *Rhinolophus malayanus* sarbecovirus genome from Laos, BANAL-52 (59), shows 96.9 % nucleotide sequence identity with SARS-CoV-2 (data not shown). However, a maximum likelihood phylogenomic tree shows that RaTG13 is the closest relative to SARS-CoV-2, with strong support (59).

Clearly, the provenance of RaTG13 is of great importance in determining the origin of the COVID-19 outbreak, whether by zoonosis, or by a lab leak event. However, a number of preprints and publications have identified potential problems with the RaTG13 raw sequence data (10) (11) (12) (13) (14) (15). In particular, database entries for the raw reads deposited on the GSA, Short Read Archive (SRA) and European Nucleotide Archive (ENA) state that the data were generated from a *R.affinis* fecal swab. In addition, the original paper describing the detection of RaTG13 RdRp (then labelled 'RaBtCov/4991'), stated that the sequence was obtained from a fecal swab (4). Likewise, a Master's thesis from the WIV describing the sequencing of the RaTG13 genome (labelled in the thesis as 'Ra4991_Yunnan'), attributed the provenance of the sample to one of 2815 anal swabs/fecal pellets collected from Yunnan Province, China from 2011-2016 (16). However, a low proportion of bacteria-related reads is indicated by a taxonomic analysis of the raw reads on the dataset's SRA webpage, and appears inconsistent with a fecal swab (10) (12) (13) (14) (15).

Here, the raw sequence reads used to generate the RaTG13 genome are analyzed in detail, using metagenomic, phylogenetic and genome mapping approaches, with the surprising finding that the genome appears to have been generated from a bat transcriptome rather than a fecal swab, indicating that RaTG13 was in live culture at the WIV. This increases the plausibility of a lab leak incident as the origin of the COVID-19 pandemic.

## Methods

### *Sequence data*

The NGS dataset used to generate the RaTG13 genome by the WIV was obtained from the GSA (accession number CRR122287). The date of collection was reported as 2013-07-24 from Yunnan, Pu'er 22.82N 100.96E (GSA Biosample accession number SAMC133252). The dataset was labelled as being generated from a fecal swab at the GSA (Experiment accession CRX097481), the SRA (accession number SRR11085797) and at the ENA (accession number SRX7724752). The GSA entry states that the QIAamp Viral RNA Mini Kit (Qiagen) was used to extract RNA and that the TruSeq

Stranded mRNA Library Preparation kit (Illumina) was used to produce the sequencing library for sequencing on the HiSeq 3000 (Illumina) platform.

A NGS dataset generated from an anal swab obtained from *R.affinis* (although the species is likely incorrect, as discussed below in Results) by Li et al. at the WIV (17) was used as a comparison. This dataset was used to generate the BtRhCoV-HKU2r (Bat Rhinolophus HKU2 coronavirus related) genome (National Center for Biotechnology Information, NCBI, accession number MN611522). The raw sequence data were obtained from the ENA (accession number SRR11085736) and were labelled as being generated from an *R.affinis* anal swab and sequenced on an HiSeq 3000 plaform. Likewise, the dataset was described as being generated from an *R.affinis* anal swab in the publication describing its genome sequence, using the QIAamp Viral RNA Mini Kit and TruSeq Library Preparation kit (Illumina) (17)

A transcriptome generated from a *R.sinicus* splenocyte cell line by the WIV was used as an additional comparison, and was obtained from the ENA (accession number SRR5819066). The protocol used to generate the dataset was not described on its ENA webpage, however it described as being sequenced using a HiSeq 2000 platform.  A NGS dataset generated by the EcoHealth Alliance from an oral swab from the bat *Miniopterus nimbae* from Zaire, and which contained Ebola, was obtained from the SRA (accession number SRR14127641). The dataset was described on its SRA webpage as being generated using the VirCapSeq target protocol (18), and sequenced using a HiSeq 4000 platform.

In this work, the four datasets will be described as the RaTG13, BtRhCoV-HKU2r anal swab, splenocyte transcriptome and Ebola oral swab datasets, respectively.

*Microbial analysis*
Metaxa2 (19) was used to identify forward reads that match small subunit (SSU) rRNA from mitochondria, bacteria and eukaryotes present in the four datasets. Phylogenetic affiliation was assigned to the lowest taxonomic rank possible from the read alignments by Metaxa2.

*Mitochondrial genome analysis*

Forward reads from the RaTG13, BtRhCoV-HKU2r anal swab and splenocyte transcriptome datasets were mapped to a variety of mitochondrial genomes corresponding to mammalian species known to have been studied at the WIV, using fastv (20).

*Mitochondrial rRNA analysis*

Some reads from the RaTG13, BtRhCoV-HKU2r anal swab and splenocyte transcriptome datasets were identified as corresponding to mitochondrial SSU rRNA using Metaxa2. These were assembled using Megahit (21). The resulting contigs were used to query the NCBI nr database using Blast (22), in order to determine its closest match. The contig generated from the RaTG13 dataset was used to make a phylogenetic tree of mitochondrial SSU rRNA genes from different *Rhinolophus* species, obtained from the NCBI (accession numbers are listed in Supplementary Table 1). Sequence alignment, model testing and phylogenetic tree construction was conducted using Mega11 (23). First, a nucleotide alignment was constructed using Muscle (24). DNA model testing was conducted and the general time reversible (GTR) model (25) was determined to be the best fit to the data using the Akaike Information Criteria (26). Then, a maximum likelihood analysis was conducted using an estimated gamma parameter and 1000 bootstrap replicates.

*Viral titer analysis*

The number of forward reads mapping to the RaTG13 genome from the RaTG13 NGS dataset were determined using fastv. Eight novel coronavirus genome sequences in addition to the BtRhCoV-HKU2r anal swab dataset were generated from NGS datasets derived from bat anal swabs from southern China by Li et al. (17). Fastv was used to determine the number of reads mapping to these nine coronavirus genomes from their respective NGS datasets.

*Nuclear genome mapping analysis*

Raw sequences from the RaTG13 and BtRhCoV-HKU2r anal swab datasets were mapped to a variety of mammalian nuclear genomes. The *Rhinolophus ferrumequinum* genome (NCBI accession number GCA_ 004115265.3) was the closest related bat genome to *R.affinis* available, and was used for mapping. In each case, the most recent assembly was used for mapping. First, the reads were trimmed and filtered using fastp (27), using polyX trimming, and filtering reads with > 5 % of reads with a quality threshold of Q < 20. Then the reads were mapped using the splicing aware mapper BBMap (https://sourceforge.net/projects/bbmap/), using the default parameters and the usemodulo option.

*Transcriptome analysis*

In order to assess the proportion of reads that mapped to gene sequences, reads from the RaTG13 dataset were mapped to a previous version of the *Rhinolophus ferrumequinum* bat nuclear genome (28) (NCBI accession number GCA_014108255.1), as a gff annotation file was not available for the most recent version of the genome assembly (GCA_ 004115265.3).

The corresponding annotation file GCA_014108255.1_mRhiFer1.p_genomic.gff was incompatible with the sam file containing mapped reads, due to differences in chromosome naming between the annotation file and corresponding genome assembly file. This was corrected by modifying the sam file using the following commands:

```
sed -i 's/ Rhinolophus ferrumequinum isolate mRhiFer1 scaffold_m29_p_[0-9][0-9]*,
whole genome shotgun sequence//g' mappingfile.sam
```

followed by

```
sed -i 's/ Rhinolophus ferrumequinum isolate mRhiFer1 mitochondrion, complete
sequence, whole genome shotgun sequence//g' mappingfile.sam
```

Quantification of the number of reads mapped to annotated genes was conducted using the bedtools multicov function (29). To do this, the sam file was first converted to a bam file and then sorted and indexed using SAMtools (30).

## Results

### *Numbers of reads matching SSU rRNA*

Only 1.8 % of forward reads in the RaTG13 dataset match SSU rRNA, in contrast with 20.7 % in the BtRhCoV-HKU2r anal swab dataset, 27.4 % in the splenocyte transcriptome dataset and 13.4 % in the Ebola oral swab dataset (Table 1). This implies it has undergone rRNA depletion during preparation, probably using the Ribo-Zero procedure which is part of the TruSeq library preparation protocol. This procedure involves enzymatic degradation of rRNA from both eukaryotes and bacteria. It is unclear if the procedure preferentially degrades rRNA from eukaryotes or bacteria, thus altering the ratio of bacterial : eukaryotic SSU rRNA sequences in the dataset.

### *Ratio of eukaryotic and bacteria SSU rRNA reads*

Consideration of the ratio of eukaryotic : bacterial SSU rRNA sequences reveals marked differences between the datasets. The ratio is 8.3 : 1 for the RaTG13 dataset, 88.7 : 1 for the splenocyte transcriptome dataset and 3.7 : 1 for the Ebola oral dataset (Table 1), indicating that eukaryotic SSU rRNA dominates these samples. However, in contrast the BtRhCoV-HKU2r anal swab dataset has a ratio of 1 : 5.4, indicating that bacterial SSU rRNA dominates this dataset, as expected with fecal material. The ratio of eukaryotic : bacterial SSU rRNAs in the RaTG13 dataset is inconsistent with that of the BtRhCoV-HKU2r anal swab dataset, and consequently appears inconsistent with fecal material.

### *Microbial SSU rRNA analysis*

Microbial taxonomic analysis provides a fingerprint that can be used to track the source of a sample by identifying taxa characteristic of the microhabitat from which it was derived (31). The results of the taxonomic analysis are displayed in Table 2. The RaTG13 dataset is dominated by *Lactococcus* spp. (64.9 % of bacterial SSU rRNA sequences). *Lactococcus* spp. are lactic acid bacteria and are not characteristic of gut microbiota. 4.9 % of bacterial SSU rRNA sequences match *Escherichia* spp., which are characteristic of the gut microbiota. However, 74.1 % of bacterial rRNA sequences also match *Escherichia* spp. in the splenocyte transcriptome dataset. The splenocyte transcriptome is dominated by members of the *Enterobacteriaceae* (90.7 % of bacterial SSU rRNAs sequences), which include members of the *Escherchia* genus. Their presence presumably reflects contamination of the culture medium, which may be characterized by a bottlenecking effect and consequent low overall microbial diversity. This is consistent with being dominated by a single bacterial family (*Enterobacteriaceae*), and with the low proportion of bacterial SSU rRNA sequences overall in the dataset (1.1 % of the total SSU rRNA sequences). Consequently, the presence of *Escherichia* spp. in the RaTG13 dataset by itself is not indicative of the presence of fecal material.

There is an absence of other bacteria characteristic of the gut in the RaTG13 dataset, in comparison with the BtRhCoV-HKU2r anal swab dataset. *Micrococcus* spp. comprise 4.5 % of bacterial SSU rRNAs sequences in the RaTG13 dataset. However, *Micrococcus* spp. are typically strict aerobes (32), and so their presence in the RaTG13 dataset appears inconsistent with a fecal swab given that the intestines are an anaerobic environment (33). *Helicobacter* spp. are diagnostic of the mammalian stomach and intestinal microbiota (34). In the BtRhCoV-HKU2r anal swab dataset 0.4 % of bacterial SSU rRNA sequences correspond to *Helicobacter* spp., but only 0.005 % in the RaTG13 dataset.

Members of the *Peptostreptococcaceae* family are anaerobes found in the human gut, soil and sediments (35). In the BtRhCoV-HKU2r anal swab dataset 21.2 % of bacterial SSU rRNA sequences correspond to members of the *Peptostreptococcaceae* family, but only 0.07 % in the RaTG13 dataset. Members of the *Lachnospiraceae* family are some of the most abundant members of the gut and intestinal microbiota in humans (36). While they comprise 6.2 % of bacterial SSU sequences in the BtRhCoV-HKU2r anal swab dataset, they only comprise 0.7 % in the RaTG13 dataset. Finally, *Clostridium* spp. are a major component of the intestinal tract (37). They comprise 47.8 % of the bacterial rRNA sequences in the anal swab dataset, but only 0.7 % of the RaTG13 dataset. These differences indicate that the microbiota of the RaTG13 sample are inconsistent with fecal material. These differences indicate that the microbiota of the RaTG13 sample are inconsistent with fecal material.

There is a possibility that the RaTG13 dataset was generated from a bat oral swab which was incorrectly labelled. Oral swabs are reported as having been collected by the WIV and the Ecohealth Alliance from 2010-2015 (38), in a joint Ecohealth Alliance-WIV NIAID grant 1R01AI 110964-01 that commenced in 2014, and in other publications by Zheng-li Shi, lead author of the Nature paper describing RaTG13 (39) (40). The issue has scientific relevance as bat coronaviruses present in oral mucosa are more likely to be transmissable via aerosols, than those that present in higher abundance in fecal swabs. Thus, determining whether RaTG13 was generated from an oral swab would provide better understanding of the emergence of SARS-CoV-2 (which is transmitted via respiratory droplets and aerosols rather than the fecal route (41)).

A taxonomic analysis of the SSU rRNA sequences from the Ebola oral swab dataset is instructive. Firstly, there is a low proportion of reads corresponding to *Lactococcus* spp. present (0.06 % of bacterial SSU rRNA sequences), in contrast with the RaTG13 dataset (64.0 % of bacterial SSU rRNA sequences). This is expected, give that lactic acid bacteria are not present in high abundance in the oral cavity. The Ebola oral swab dataset has a low proportion of *Escherichia* spp. (0.003 % of bacterial SSU rRNA sequences), in contrast to the RaTG13 dataset (4.9 % of bacterial SSU rRNA sequences). *Escherichia* spp. are not expected in the oral cavity as they are intestinal bacteria, and coprophagy is unreported in bats, consequently their presence in the RaTG13 dataset is inconsistent with it being generated from an oral swab.

Members of the *Pasteurellaceae* family are mostly commensals living on mucosal surfaces, particularly in the upper respiratory tract (42). The Ebola oral swab dataset is dominated by members of the *Pasteurellaceae* family (51.3 % of bacterial SSU rRNA sequences), in contrast with only 0.03 % in the RaTG13 dataset. The *Haemophilus* genus is part of the *Pasteurellaceae* family and *Haemophilus* spp. are characteristic of the upper respiratory tract (43). They comprise 4.4 % of bacterial SSU rRNA sequences in the Ebola oral swab dataset, but only 0.01 % from the RaTG13 dataset. The near absence of members of the *Pasteurellaceae* family in the RaTG13 is inconsistent with it being derived from an oral swab. Members of the *Gemella* genus are characteristic of the oral microbiota in humans (44). *Gemella* spp. comprise 0.3 % of the bacterial SSU rRNA sequences in the Ebola oral swab dataset, but are completely absent in the RaTG13 dataset.

While some sequences in the BtRhCoV-HKU2r anal swab dataset might be expected to originate from the bat insectivorous diet (45), only a few arthropod nuclear rRNA sequences were observed in the BtRhCoV-HKU2r, RaTG13 anal swab and splenocyte transcriptome datasets (0.01 %, 0.02 % and 0.01 % of eukaryotic SSU rRNA sequences, respectively). However, the Ebola oral swab dataset has substantially more (0.4 % of eukaryotic SSU rRNA sequences). This is consistent with the insectivorous diet of *M.nimbae*, the bat species from which the oral swab was taken. *Rhinolophus* spp. are also insectivorous, and so the lower relative proportion of arthropod SSU rRNA sequences in the RaTG13 sample is an additional inconsistency with an oral swab. The observations listed above indicate the differences between Ebola oral swab microbiota, which are consistent with an oral microhabitat, with the microbiota present in the RaTG13 sample. These data indicate that the RaTG13 sample was not derived from an oral swab.

*Mitochondrial analysis*

Reads from the RaTG13, BtRhCoV-HKU2r anal swab and splenocyte transcriptome datasets were mapped onto a range of mammalian mitochondrial genomes (Table 3). Reads from the RaTG13 dataset mapped most efficiently to the *R.affinis* mitochondrial genome with 75335 reads mapping with 97.2 % coverage. 18017 reads mapped with 40.4 % coverage to the *R.sinicus* mitochondrial genome. This implies that the sample originated from *R.affinis* or a *Rhinolophus* species more closely related to *R.affinis* than *R.sinicus*.

The low proportion of total reads mapping to the *R.affinis* mitochondrial genome (0.3 %) suggests the RaTG13 sample was subjected to DNase treatment during preparation, which is an optional step in the QIAamp Viral RNA Mini Kit protocol. This is consistent with the observation that the majority of reads that map to the *Rhinolophus ferrumequinum* genome map to annotated protein coding genes (discussed below), implying little host nuclear DNA was present in the sample.

Reads from the BtRhCoV-HKU2r anal swab dataset mapped most efficiently to the *R.sinicus* mitochondrial genome, with 29.8 % coverage and 10019 reads mapping, in contrast to the *R.affinis* mitochondrial genome, which mapped with 14.9 % coverage and 6278 reads mapping. This indicates that the sample was derived from *R.sinicus* or a more closely related *Rhinolophus* species than *R.affinis*, and is consistent with the phylogenetic analysis below. This contradicts the description of the sample as being derived from *R.affinis*.

Lastly, reads from the splenocyte transcriptome dataset mapped most efficiently to the *R.sinicus* mitochondrial genome, with 94.5 % coverage and 170591 reads mapping, in contrast to the *R.affinis* mitochondrial genome, which mapped with 32.2 % coverage and 88220 reads mapping. This indicates that the sample was derived from *R.sinicus* or a more closely related *Rhinolophus* species than *R.affinis*.

*Mitochondrial rRNA analysis*

While mapping to the mitochondrial genome gives a convincing indication of the general phylogenetic affinities of the NGS datasets, mitochondrial SSU rRNA confers more precision. A 1139 bp contig generated by Megahit from SSU rRNA sequences extracted from the RaTG13 dataset using Metaxa2 was found to match *R.affinis* mitochondrial mitochondrial SSU rRNA (NCBI accession number MT845219) with 98.7 % sequence identity, with 8 mismatches (Figure 1). A maximum likelihood phylogenetic tree indicates that the RaTG13 contig was most closely related to *R.affinis* mitochondrial SSU rRNA, compared to other *Rhinolophus* species for which full length mitochondrial SSU rRNA sequences were available (Figure 2).

Mitochondrial SSU rRNA sequences generated by Metaxa2 from the BtRhCoV-HKU2r anal swab dataset were likewise assembled using Megahit. A 960 bp contig aligned to *Rhinolophus sinicus sinicus* mitochondrial SSU rRNA (NCBI accession number KP257597.1), with only one mismatch. This is surprising given that the anal swab sample is described as having been obtained from *R.affinis* (17) (supplementary file msphere.00807-19-st002.xlsx), as is the BtRhCoV-HKU2r coronavirus genome sequence (NCBI accession number MN611522). The result is consistent with the mitochondrial mapping results described above.

Finally, mitochondrial rRNA sequences generated by Metaxa2 from the splenocyte transcriptome dataset were assembled using Megahit. This produced a contig of 943 bp bp, which perfectly aligned to *R.sinicus sinicus* mitochondrial SSU rRNA, with no mismatches. This is consistent with the database description of the dataset as having been derived from a *R.sinicus* cell line.

The almost perfect match of the BtRhCoV-HKU2r anal swab dataset and the perfect match of the splenocyte transcriptome dataset to *R.sinicus sinicus* mitochondrial SSU rRNA demonstrates the accuracy of the methodology in generating high quality mitochondrial SSU rRNA sequences from bat NGS RNA datasets, whether anal swab or cell line. The number of mismatches displayed by mitochondrial SSU rRNA sequence from the RaTG13 to *R.affinis* mitochondrial SSU rRNA is interesting therefore, as these are unlikely to have arisen as the result of sequencing or assembly errors.

This is consistent with the observation that an 866 bp contig generated from mitochondrial SSU rRNA sequences extracted from the reverse read dataset (CRR122287_f1) perfectly matched the 1139 bp contig generated from the forward read dataset (CRR122287_r2), where they overlapped (data not shown). If the mismatches of the 1139 bp with the *R.affinis* mitochondrial SSU rRNA reference sequence were due to sequencing or assembly errors, they would not be observed in the 866 bp contig. In addition, when the 1139 bp contig sequence is aligned with the mitochondrial SSU rRNA sequences of other *Rhinolophus* species included in the phylogenetic analysis, only 3 mismatches with the *R.affinis* sequence are unique, while the other 5 mismatches are also observed in other species in the alignment (Supplementary Figure 1).

The eight mismatches of the 1139 bp contig to *R.affinis* mitochondrial SSU rRNA (derived from the subspecies *himalayanus*, sampled from Anhui province (46)) implies that the dataset was derived from a genetically distinct population/subspecies of *R.affinis*, or a closely related species. This is consistent with the observation that the *R.affinis* taxon has nine subspecies with marked morphological and echolocation differences, and might actually represent a species complex (47). In addition, *Rhinolophus stheno*, a species closely related to *R.affinis* (48) (49), was identified as being present in the mine in Mojiang in 2015 (50). Unfortunately, no mitochondrial SSU rRNA sequence is currently available from this species.

*Evidence of sequence contamination*

Megahit-assembled contigs corresponding to plant nuclear SSU rRNA sequences were recovered from both the RaTG13 and BtRhCoV-HKU2r anal swab datasets. A 250 bp contig was recovered from the RaTG13 dataset that showed 100 % identity to *Gossypium hirsutum* (cotton) nuclear SSU rRNA. A 366 bp contig was recovered from the anal swab dataset that showed 100 % identity to *Zea mays* (maize) nuclear SSU rRNA, while a 390 bp contig showed 98.5 % identity to *Arabis alpine* (alpine rock cress) nuclear SSU rRNA, with 6 mismatches (data not shown).

*Rhinolophus* bats are insectivorous and so it is difficult to see how plant material could have entered the samples. Apparent contamination of agricultural NGS datasets generated by Huazhong Agricultural University (Wuhan) by coronavirus sequences have been previously reported, and may originate from index hopping, or cross contamination of sequencing reagents (51). A Hiseq Illumina sequencing machine ST-J00123 owned by Novogene appears to have serviced institutions from the Wuhan area, including both Huazhong Agricultural University and the WIV (52). The sequencing of plant samples on the same machine as coronavirus samples may account for contamination of the WIV RaTG13 raw reads.

*Viral titer comparison*

A comparison was made of the number of coronavirus reads present in the RaTG13 sample, BtRhCoV-HKU2r anal swab sample, and eight additional bat anal swab samples generated by the WIV by Li et al. (17) (Table 4). These data shows that the viral titer in the RaTG13 sample was relatively low ($7.2 \times 10^{-5}$ of total reads map to the coronavirus genome), compared to the nine anal swab samples generated by Li et al. (which includes the BtRhCoV-HKU2r anal swab sample), which ranged from $3.0 \times 10^{-5}$ to $4.9 \times 10^{-2}$ of total reads mapping to the respective coronavirus genomes. Unfortunately, there are no coronavirus datasets generated from cell lines by the WIV available for comparison. Finally, it was found that raw reads generated from *Rhinolophus larvatus* (SRA accession number SRR11085733) mapped to the BtHiCoV-CHB25 genome, and not *Hipposideros pomona* as reported in the supplementary file msphere.00807-19-st002.xlsx (17).

*Genome mapping analysis*

In order to identify the origin of the bulk of reads in the RaTG13 dataset, they were mapped to a variety of mammalian genomes, corresponding to species for which cell lines were known to be in use at the WIV, as well as *Rhinolophus ferrumequinum*, which is the bat genome most closely related to *R.affinis* available (Table 5). The results show that the reads most efficiently map to the *R.ferrumequinum* genome, with 87.5 % of reads mapping. An even higher percentage of reads would be expected to map to the exact *Rhinolophus* species used to generate the RaTG13 sample (*R.affinis* or closely related species, as identified in the phylogenetic analysis above).

The high percentage of reads mapping to *R.ferrumequinum* is inconsistent with a fecal swab, which is expected to have a majority of reads mapping to bacterial sequences. This is because fecal material is typically dominated by bacteria, with only a small amount of host nucleic acid present (53). Consistent with this expectation, only 2.6 % of the reads from the BtRhCoV-HKU2r anal swab sample mapped to the *R.ferrumequinum* genome (Table 5). In addition, the results appear inconsistent with an oral swab. This is because only 27.9 % of reads from the Ebola swab sample map to the *Miniopterus natalensis* genome (NCBI accession number GCF_001595765.1), which was the most closely related bat genome to *M.nimbae* available. However, only the forward reads from the NGS dataset were used for the mapping, as the reverse reads were not available. In addition, the RNA purification method used is unclear from the NCBI sample webpage. These two factors means that the Ebola oral swab mapping results may not be directly comparable to the RaTG13- *R.ferrumequinum* mapping results.

*Transcriptome analysis*

Further analysis of the RaTG13 reads mapped to the *R.ferrumequinum* genome shows that 62.1 % of mapped reads map to protein coding genes. 92.1 % of protein coding genes have at least one read that maps to it. These data confirm that the RaTG13 sample represents a transcriptome. In addition, the result indicates that the sample did not have large amounts of DNA present as this would lead to mapping to parts of the genome that do not code for protein coding genes, which is the large majority of the bat genome. This supports the mitochondrial genome mapping results that the sample was subjected to DNase treatment, which is an optional step in the QIAamp Viral RNA Mini Kit.

## Discussion

The data presented here indicate that the RaTG13 genome was not generated from a bat fecal swab, but rather a *Rhinolophus* sp. cell line or tissue. Given that the original RaBtCov/4991 RdRp sequence fragment was described as having been generated from a *R.affinis* fecal swab (4), then the chain of events leading from that original sample to the sample used to generate the genome sequence is unclear.

As far as the author is aware, no live animals were reported as being captured in the collecting expeditions to Mojiang between August 2012 to July 2013 and there is no reported precedent for virus isolation from bat tissue at the WIV. If the sample was derived from a dead bat, it is hard to understand how the sample became depleted, as stated by Zheng-liShi (54), given that tissue would have yielded substantial RNA. The presence of a substantial proportion of reads corresponding to the *Escherichia* spp. and *Lactococcus* spp. in the NGS reads (Table 2) would also be hard to understand.

There are two examples of mislabelling of coronavirus samples by researchers at the WIV identified here. The BtRhCoV-HKU2r anal swab sample appears to have been derived from *R.sinicus sinicus* and not *R.affinis* as described. In addition, the BtHiCoV-CHB25 anal swab sample was derived from *R.larvatus* and not *H.pomona* as reported. These observations indicate that sample collection and processing were error prone, lending some credence to a lab leak scenario for the origin of the COVID-19. Given that these samples may contain potential pandemic pathogens (PPPs), this is of great concern. In particular, it is of note that in a Master's thesis by Yu Ping, supervised by Zheng-li Shi and Cui Jie, work on the Ra4991_Yunnan (RaTG13) sample and other bat samples is described as being conducted in a 'BSL-2 cabinet', but it is unclear if the cabinet was situated in a BSL-2 lab or a regular lab (16). Bat coronaviruses are PPPs and so should be handled in a BSL-3 lab (55). A further curiosity is that neither Yu Ping or Cui Jie are listed as authors in the Nature paper describing RaTG13 (3).

The data imply that RaTG13 may have been in live culture at the WIV since before June 2017, which is when the Illumina sequencing of the RaTG13 (RaBtCOv/4991) sample appears to have begun (information generated by Francisco de Asis de Ribera). This implies that isolation of RaTG13 was successfully conducted on the original sample collected from the Mojiang mine after its collection in July 2013 and before June 2017. It is unclear if RaTG13 is currently in live culture at the WIV. Of high relevance to the work described here is the statement on page 119 in the recently released EcoHealth-WIV grant 1R01AI 110964-01 that the WIV had successfully isolated coronaviruses using bat cell lines:

'*We have developed primary cell lines and transformed cell lines from 9 bat species* using kidney, spleen, heart, brain and intestine. *We have used these for virus isolation*, infection assays and receptor molecule gene cloning.' (italics the author's)

**Conclusion**

In the Nature paper describing RaTG13 (3), and its Addendum (6), there is no specific statement that the RaTG13 raw reads were generated from a fecal swab. However, the dataset is labelled as such at the GSA, SRA and ENA, the original RaBtCoV /4991 sample from the Mojiang mine is described as having been a fecal swab (4), and the Master's thesis describing the genome sequencing of RaTG13 describes it as having been generated from an anal swab/fecal pellet (16).

Consequently, the Nature paper describing RaTG13 and its database entries should be amended to state the true provenance of RaTG13. Pertinent to this is the observation that serial passaging during stock maintenance typically causes SARS-CoV-2 to diverge genetically, given that is rapidly adapts to the cell culture conditions (56). This means that the reported RaTG13 genome sequence would be expected to have picked up mutations during its laboratory sojourn, from its date of isolation. This is consistent with the observation that the genome has an excess of synonymous mutations compared to the closely related *Rhinolophus malayanus* RmYN02 coronavirus (which was sampled in 2019 (57)), in relation to SARS-CoV-2 (58). This could be interpreted as being the result of evolutionary change occurring from its 2013 collection date, which could only happen if the virus was in culture (as noted in a preprint of (58) at https://www.biorxiv.org/content/10.1101/2020.04.20.052019v1).

The observations outlined here have important implications regarding the origin of the COVID-19 pandemic. Zheng-li Shi, lead author of the Nature paper, has stated that RaTG13 was not in live culture at the WIV, as follows: “I would like to emphasize that we have only the genome sequence and didn’t isolate this virus” and “…we did not do virus isolation and other studies on it” (54). The data presented here suggest otherwise, and point to the possibility of additional coronaviruses closely related to SARS-CoV-2 in culture at the WIV that have not yet been disclosed. This conclusion increases the plausibility of the lab leak scenario for the origin of COVID-19.

**Acknowledgements**

The investigation into the origin of the COVID-19 pandemic has represented a paradigm shift for how forensic and epidemiological investigations are conducted, with decentralized online groups of investigators making significant contributions. This work is the product of discussions with numerous investigators based on Twitter, some anonymous, several associated with the Decentralized Radical Autonomous Search Team Investigating COVID-19 (DRASTIC) (https://drasticresearch.org/), and others independent. The author has made widespread use of material generated by online investigators, including information generated by @BillyBostickson, @TheSeeker268, @franciscodeasis, @Daoyu15, @pathogenetics, @mrandersoninneo and @Ayjchan. The Master’s thesis of Yu Ping was identified and translated by @TheSeeker268 and @franciscodeasis. The EcoHealth-WIV NIAID grant 1R01AI 110964-01 was made available by an FOIA made by The Intercept (theintercept.com). In particular, the author would like to thank @Florin_Uncovers for his tireless (59) (59) persistence and curiosity: this work is the direct result of our discussions together.

**Figure 1** Alignment of mitochondrial SSU rRNA sequence assembled from the RaTG13 sample with the *R.affinis* reference sequence

```
R.affinis        ------------------------------------------------------------
RaTG13-sample    CTGTTAATGTAGCTTAATCAACCAAAGCAAGGCACTGAAAATGCCTAGATGAGTATTAAT


R.affinis        ------------CATAGGCTTGGTCCTGGCCTTTCTGTTGGTTCTAGGTAAAACTACACG
RaTG13-sample    ACTCCATAAACACATAGGCTTGGTCCTGGCCTTTCTGTTGGTTTTAGGTAAAACTACACA
                             *******************************.***************.

R.affinis        TGCAAGTATCTGCA-CCCAGTGAGAATGCCCTCTAAATCACACCTGATTAAAAGGAGCGG
RaTG13-sample    TGCAAGTATCCGCACCCCAGTGAGAATGCCCTCTAAATCACGCCTGATTAAAAGGAGCGG
                 **********.*** **************************.******************

R.affinis        GCATCAAGCACACTACAAAGTAGCTCATGACGCCTTGCTTAACCACGCCCCCACGGGAAA
RaTG13-sample    GCATCAAGCACACTACAAAGTAGCTCATGACGCCTTGCTTAACCACGCCCCCACGGGAAA
                 ************************************************************

R.affinis        CAGCAGTGATAAAAATTAAGCCATGAACGAAAGTTCGACTAAGTTATACCTACTCCTTAG
RaTG13-sample    CAGCAGTGATAAAAATTAAGCCATGAACGAAAGTTCGACTAAGTTATACCTACTCCTTAG
                 ************************************************************

R.affinis        GGTTGGTAAATTTCGTGCCAGCCACCGCGGTCACACGATTAACCCAAATTAACAGAAACA
RaTG13-sample    GGTTGGTAAATTTCGTGCCAGCCACCGCGGTCACACGATTAACCCAAATTAACAGAAACA
                 ************************************************************

R.affinis        CGGCGTAAAGCGTGTTTAAGAATAC-AAAAAAAATAAAGTTAAATTCTAGCTAAGCTGTA
RaTG13-sample    CGGCGTAAAGCGTGTTTAAGAATACAAAAAAAAATAAAGTTAAATTCTAGCTAAGCTGTA
                 ************************* **********************************

R.affinis        AAAAGCCATAGCTAAAATAAAAATAGACTACGAAAGTGACTTTACAAATTCTGAATACAC
RaTG13-sample    AAAAGCCATAGCTAAAATAAAAATAGACTACGAAAGTGACTTTACAAATTCTGAATACAC
                 ************************************************************

R.affinis        GATAGCTAAGACCCAAACTGGGATTAGATACCCCACTATGCTTAGCCCTAAACCTAAACA
RaTG13-sample    GATAGCTAAGACCCAAACTGGGATTAGATACCCCACTATGCTTAGCCCTAAACCTAAACA
                 ************************************************************

R.affinis        ATCAACACAACAACATTATTCGCCAGAGTACTACTAGCAACAGCTTAAAACTCAAAGGAC
RaTG13-sample    ATCAACACAACAACATTATTCGCCAGAGTACTACTAGCAACAGCTTAAAACTCAAAGGAC
                 ************************************************************

R.affinis        TTGGCGGTGCTTCATACCCCTCTAGAGGAGCCTGTCCTATAATCGATAAACCCCGATAGA
RaTG13-sample    TTGGCGGTGCTTCATACCCCTCTAGAGGAGCCTGTCCTATAATCGATAAACCCCGATAAA
                 **********************************************************.*

R.affinis        CCTCACCAGCTCTTGCCAATTCAGCCTATATACCGCCATCCTCAGCAAACCCTAAAAAGG
RaTG13-sample    CCTCACCAGCTCTTGCCAATTCAGCCTATATACCGCCATCCTCAGCAAACCCTAAAAAGG
                 ************************************************************

R.affinis        AACTGCAGTAAGCACAAACATTAGACATAAAAACGTTAGGTCAAGGTGTAGCCTATGAGC
RaTG13-sample    AACTGCAGTAAGCACAAACATTAGACATAAAAACGTTAGGTCAAGGTGTAGCCTATGAGC
                 ************************************************************
```

```
R.affinis        TGGGAAGAGATGGGCTACATTTTCTTCTCAAAGAACATTTAAAACTACATACGGAAGTTC
RaTG13-sample    TGGGAAGAGATGGGCTACATTTTCTTCTCAAAGAACATTTAAAACTACATACGGAAGTTC
                 ************************************************************

R.affinis        TCATGAAATAGAGAGCGGAAGGTGGATTTAGTAGTAAATCAAGAACAAAGAGCTTGGTTG
RaTG13-sample    TCATGAAATAGAGAACGGAAGGTGGATTTAGTAGTAAATCAAGAACAAAGAGCTTGGTTG
                 **************.*********************************************

R.affinis        AATTAGGCCATGAAGCACGCACACACCGCCCGTCACCCTCCTCAAATATGAAGGTAATAC
RaTG13-sample    AATTAGGCCATGAAGCACGCACACACCGCCCGTCACCCTCCTCAAATATGAAGGTAATGC
                 **********************************************************.*

R.affinis        CCAAACCTATTACCACACACCCACAATATGAGAGGAGATAAGTCGTAACAAGGTAAGCGT
RaTG13-sample    CCAAACCTATTACCACACACCCACAATATGAGAGGAGATAAGTCGTAACAAGGTAAGCGT
                 ************************************************************

R.affinis        ACTGGAAAGTGCGCTTGGATACAT------------------------------------
RaTG13-sample    ACTGGAAAGTGCGCTTGGATACATCAAAGTGTAGCTTAAGCCAAAAGCACCTGGCTTACA
                 ************************

R.affinis        -----------------------------------------------------------
RaTG13-sample    CCCAGGAGACTTCACGTACAATGAACGCTTTGAACAAGTACTAGCCCAACCACAACCCA
```

**Figure 2** Maximum likelihood phylogenetic tree of mitochondrial SSU rRNA generated from the RaTG13 sample and *Rhinolophus* species

The tree was constructed as described in Methods, accession numbers for the *Rhinolophus* mitochondrial SSU rRNA sequences used are in Supplementary Table 1. Numbers at nodes represent percentages from 1000 bootstrap replicates.

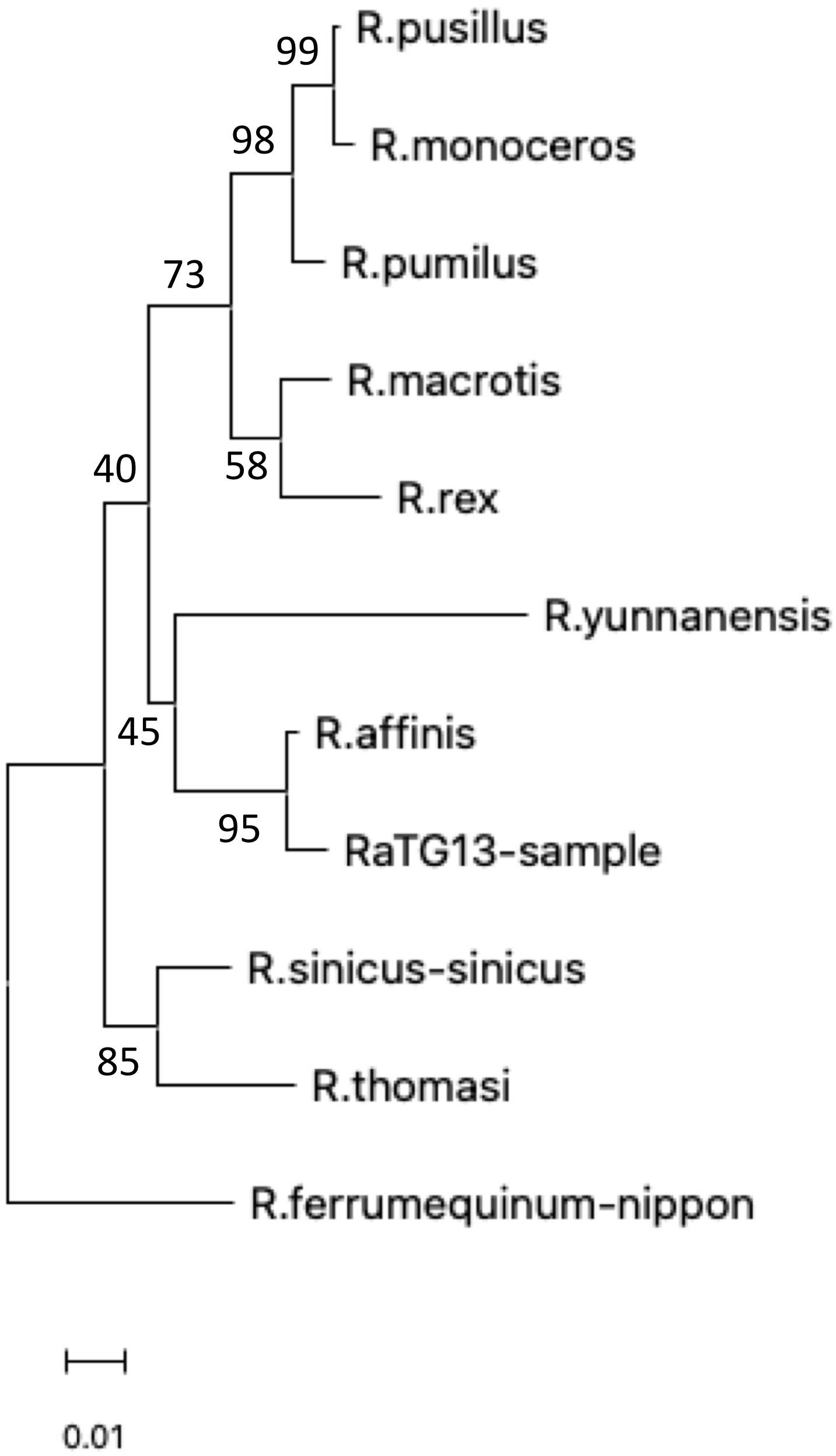

**Table 1** Numbers of eukaryotic and bacterial SSU rRNA reads present in the NGS datasets

SSU rRNA reads were identified in each dataset as described in Methods.

| Sample | Total number of (forward) reads in dataset | Total number of rRNA sequences (% of total reads in brackets) | Number of bacterial rRNA sequences (% of total rRNA sequences in brackets) | Number of eukaryotic rRNA sequences (% of total rRNA sequences in brackets) | Ratio of eukaryotic:bacterial rRNA sequences |
|---|---|---|---|---|---|
| RaTG13 | 11604666 | 208776 (1.8 %) | 21548 (10.3 %) | 178804 (85.6 %) | 8.3 : 1 |
| BtRhCoV-HKU2r anal swab | 11924182 | 2470567 (20.7 %) | 2085824 (84.4 %) | 384023 (15.5 %) | 1 : 5.4 |
| Splenocyte transcriptome | 4764112 | 1306781 (27.4 %) | 13959 (1.1 %) | 1238388 (94.8 %) | 88.7 : 1 |
| Ebola oral swab | 1000000 | 1341026 (13.4 %) | 283350 (21.1 %) | 1050512 (78.3 %) | 3.7 : 1 |

**Table 2** Taxonomic analysis of SSU rRNA sequences present in the datasets
SSU rRNA sequences present in the NGS datasets were identified as described in Methods. The percentages represent the proportion of sequences corresponding to the respective taxonomic group in the different samples (number of sequences in brackets). For bacteria, this was calculated as the proportion of bacterial SSU rRNA reads and for eukaroytes, this was calculated as the proportion of eukaryotic SSU rRNA reads.

| Taxonomic group | RaTG13 sample | BtRhCoV-HKU2r anal swab | Splenocyte transcriptome | Ebola oral swab |
|---|---|---|---|---|
| **Bacteria** | | | | |
| *Enterobacteriaceae* | 18.1 % (3891) | 7.4 % (154498) | 90.7 % (12667) | 2.4 % (6860) |
| *Enterobacteriaceae*, *Escherichia* | 4.9 % (1046) | 4.8 % (100567) | 74.1 % (10340) | 0.003 % (9) |
| *Mycoplasma* | 0.07 % (15) | 0.04 % (869) | 0.1 % (18) | 0.005 % (14) |
| *Helicobacter* | 0.005 % (1) | 0.4 % (8490) | 0 % (0) | 0 % (0) |
| *Bacillus* | 0.004 % (8) | 0.4 % (8214) | 0.01 % (1) | 5.2 % (14662) |
| *Peptostreptococcaceae* | 0.07 % (16) | 21.2 % (442050) | 0 % (0) | 0.0007 % (2) |
| *Enterococcus* | 6.7 % (1453) | 6.3 % (132235) | 0.2 % (30) | 2.4 % (6828) |
| *Lachnospiraceae* | 0.7 % (146) | 6.2 % (128630) | 0.01 % (1) | 0.08 % (221) |
| *Clostridium* | 0.7 % (141) | 47.8 % (997409) | 0 % (0) | 0.01 % (31) |
| *Lactococcus* | 64.0 % (13780) | 0.07 % (1532) | 0 % (0) | 0.06 % (162) |
| *Lactobacillus* | 0.02 % (5) | 0.0009 % (18) | 0.01 % (1) | 0.005 % (13) |
| *Micrococcus* | 4.5 % (960) | 0 % (0) | 0.08 % (11) | 0 % (0) |
| *Pasteurellaceae* | 0.03 % (7) | 3.1 % (64468) | 0 % (0) | 51.3 % (145394) |
| *Pasteurellaceae, Haemophilus* | 0.01 % (2) | 0.02 % (511) | 0 % (0) | 4.4 % (12383) |
| *Gemella* | 0 % (0) | 0.005 % (106) | 0.02 % (3) | 0.3 % (742) |
| | | | | |
| **Eukaryota** | | | | |
| *Arthropoda* | 0.02 % (35) | 0.01 % (36) | 0.01 % (157) | 0.4 % (4005) |
| *Arthropoda*, Insecta | 0.02 % (29) | 0.008 % (30) | 0.009 % (113) | 0.1 % (1348) |
| *Fungi* | 0.2 % (302) | 0.002 % (6) | 0.02 % (261) | 0.02 % (186) |
| *Viridiplantae* | 0.1 % (214) | 0.05 % (198) | 0.001 % (13) | 0.002 % (22) |

**Table 3** Mapping of NGS sample reads to mammalian mitochondrial genomes

NGS sample reads were mapped to a series of mammalian mitochondrial genomes as described in Methods.

| Species | Mitochondrial genome NCBI accession number | Percent of mitochondrial genome covered (number of reads mapped in brackets) | | |
|---|---|---|---|---|
| | | RaTG13 | BtRhCoV-HKU2r anal swab | Splenocyte transcriptome |
| *R.affinis* | NC_053269.1 | 97.2 % (75335) | 14.9 % (6278) | 32.2 % (88220) |
| *R.sinicus* | KP257597.1 | 40.4 % (18017) | 29.8 % (10019) | 94.5 % (170591) |
| Mouse | NC_005089.1 | 6.3 % (111) | 1.6 % (18) | 6.0 % (1755) |
| Human | NC_012920.1 | 3.6 % (26) | 9.4 % (23) | 40.5 % (91) |
| Black foot ferret (*Mustela nigripes*) | NC_024942.1 | 6.6 % (1537) | 3.0 % (61) | 5.2 % (1383) |
| Malaysian pangolin (*Manis javanica*) | NC_026781.1 | 4.9 % (88) | 2.2 % (32) | 3.5 % (1254) |
| Pig | NC_012095.1 | 6.6 % (1606) | 4.2 % (155) | 5.2 % (2238) |
| Rabbit (*Oryctolagus cuniculus*) | NC_001913.1 | 5.1 % (92) | 1.4 % (16) | 2.7 % (1529) |
| Asian Palm civet | MG200264.1 | 5.6 % (1836) | 4.3 % (185) | 4.4 % (5258) |
| Chinese tree shrew (*Tupaia chinensis*) | AF217811 | 4.2 % (655) | 2.7 % (14) | 2.9 % (4117) |

**Table 4** Number of coronavirus sequences present in the RaTG13 sample, and anal swab samples from the WIV

Nine coronavirus genomes were generated by the WIV from anal swabs (17). The raw (forward) reads were mapped to the respective coronavirus genomes as described in Methods. The reads from the RaTG13 sample were mapped to the RaTG13 genome in comparison.

| Sample dataset SRA accession number (number of reads in brackets) | Coronavirus genome (NCBI accession number in brackets) | Number of reads mapped to respective coronavirus genome | Proportion of reads mapping to coronavirus compared to total number of reads |
|---|---|---|---|
| SRR11085797 (23209332) | RaTG13 (MN996532) | 1669 | $7.2 \times 10^{-5}$ |
| SRR11085736 (23848364) | BtRhCoV-HKU2r (MN611522) | 886 | $3.7 \times 10^{-5}$ |
| SRR11085735 (8032494) | BtHpCoV-HKU10-related (MN611523) | 7030 | $8.8 \times 10^{-4}$ |
| SRR11085733 (*R.larvatus*) (27083324) | BtHiCoV-CHB25 (MN611525) | 1035522 | $3.8 \times 10^{-2}$ |
| SRR11085741 (24828142) | BtRaCoV-229E-related (MN611517) | 99776 | $4.0 \times 10^{-3}$ |
| SRR11085734 (19171950) | BtMiCoV-1-related (MN611524) | 581 | $3.0 \times 10^{-5}$ |
| SRR11085740 (19562848) | BtMiCoV-HKU8-related (MN611518) | 2817 | $1.4 \times 10^{-4}$ |
| SRR11085737 (23088962) | BtScCoV-512-related (MN611521) | 142646 | $6.2 \times 10^{-3}$ |
| SRR11085738 (29134128) | BtPiCoV-HKU5-related (MN611520) | 1437700 | $4.9 \times 10^{-2}$ |
| SRR11085739 (9589348) | BtTyCoV-HKU4-related (MN611519) | 2778 | $2.9 \times 10^{-4}$ |

**Table 5** Mapping of sample reads to mammalian nuclear genomes

| Species | Genome assembly NCBI accession number | % of RaTG13 sample reads mapped to genome | % of BtRhCoV-HKU2r anal swab sample reads mapped to genome |
|---|---|---|---|
| Greater horseshoe bat (*R.ferrumequinum*) | GCA_ 004115265.3 | 87.5 % | 2.6 % |
| Human | GCA_000001405.28 | 64.3 % | 7.4 % |
| Mouse | GCA_000001635.9 | 62.2 % | 6.6 % |
| Green monkey | GCA_000409795.2 | 51.5 % | 7.5 % |
| Pig | GCA_000003025.6 Sscrofa11.1 | 62.5 % | 7.4 % |

**Supplementary Materials**

**Supplementary Figure 1** Multiple sequence alignment of mitochondrial SSU rRNA sequences from *Rhinolophus* species

An alignment was constructed using a 1139 bp contig generated from the RaTG13 dataset as described in Methods, and mitochondrial SSU rRNA sequences from *Rhinolophus* species (accession numbers in Supplementary Table 1). Yellow indicates a unique polymorphism present in the RaTG13 sequence, while blue indicates a mismatch between this sequence and the *R.affinis* sequence, but that is consistent with other *Rhinolophus* sp. sequences present in the alignment. Consequently, the nucleotides highlighted in blue are not unique to the RaTG13 sequence. The alignment was constructed using Muscle.

```
R.yunnanensis              ------------------------------------------------------------
R.ferrumequinum-nippon     ------------------------------------------------------------
R.pumilus                  ------------------------------------------------------------
R.pusillus                 ------------------------------------------------------------
R.monoceros                ------------------------------------------------------------
R.macrotis                 ------------------------------------------------------------
R.rex                      ------------------------------------------------------------
R.affinis                  ------------------------------------------------------------
RaTG13-sample              CTGTTAATGTAGCTTAATCAACCAAAGCAAGGCACTGAAAATGCCTAGATGAGTATTAAT
R.sinicus-sinicus          ------------------------------------------------------------
R.thomasi                  ------------------------------------------------------------


R.yunnanensis              ------------CATAGGCTTGGTCCTGGCCTTTCTGTTGGTTCTAGGTAAAACTACACA
R.ferrumequinum-nippon     ------------CATAGGCTTGGTCCTGGCCTTTCTGTTAGTTCTAGGTAAAACTACACA
R.pumilus                  ------------CATAGGCTTGGTCCTGGCCTTTCTGTTGGTTCTAGGTAAAACTACACA
R.pusillus                 ------------CATAGGCTTGGTCCTGGCCTTTCTGTTGGTTCTAGGTAAAACTACACA
R.monoceros                ------------CATAGGCTTGGTCCTGGCCTTTCTGTTGGTTCTAGGTAAAACTACACA
R.macrotis                 ------------CATAGGCTTGGTCCTGGCCTTTCTGTTGGTTCTAGGTAAAACTACACA
R.rex                      ------------CATAGGCTTGGTCCTGGCCTTTCTGTTGGTTCTAGGTAAAACTACACA
R.affinis                  ------------CATAGGCTTGGTCCTGGCCTTTCTGTTGGTTCTAGGTAAAACTACACG
RaTG13-sample              ACTCCATAAACACATAGGCTTGGTCCTGGCCTTTCTGTTGGTTTTAGGTAAAACTACACA
R.sinicus-sinicus          ------------CATAGGCTTGGTCCTGGCCTTTCTGTTGGTTCTAGGTAAAACTACACA
R.thomasi                  ------------CATAGGCTTGGTCCTGGCCTTTCTGTTGGTTCTAGGTAAAACTACACA
                                       *************************** *** ***************


R.yunnanensis              TGCAAGTATCCGCACTCCAGTGAGAATGCCCTCTAAATCACACCTGATTAAAAGGAGCGG
R.ferrumequinum-nippon     TGCAAGTATCCGCACCCCAGTGAGAATGCCCTCTAAATCACACCTGATTAAAAGGAGCGG
R.pumilus                  TGCAAGTATCCGCACTCCAGTGAGAATGCCCTCTAAATCATATCTGATTAAAAGGAGCGG
R.pusillus                 TGCAAGTATCCGCACTCCAGTGAGAATGCCCTCTAAATCATACCTGATTAAAAGGAGCGG
R.monoceros                TGCAAGTATCCGCACTCCAGTGAGAATGCCCTCTAAATCATACCTGATTAAAAGGAGCGG
R.macrotis                 TGCAAGTATCCGCACTCCAGTGAGAATGCCCTCTAAATCACGCCTGATTAAAAGGAGCGG
R.rex                      TGCAAGTATCCGCACTCCAGTGAGAATGCCCTCTAAATCATGCCTGATTAAAAGGAGCGG
R.affinis                  TGCAAGTATCTGCAC-CCAGTGAGAATGCCCTCTAAATCACACCTGATTAAAAGGAGCGG
RaTG13-sample              TGCAAGTATCCGCACCCCAGTGAGAATGCCCTCTAAATCACGCCTGATTAAAAGGAGCGG
R.sinicus-sinicus          TGCAAGTATCCGCGCCCCAGTGAGAATGCCCTCTAAATCATACCTGATTAAAAGGAGCGG
R.thomasi                  TGCAAGTATCCGCACTCCAGTGAGAATGTCCTCTAAATCACACCTGATTAAAAGGAGCGG
                           ********** ** * ************ ***********   *****************


R.yunnanensis              GCATCAAGCGCACTACAAAGTAGCTCATAACGCCTTGCTTAACCACACCCCCACGGGAAA
R.ferrumequinum-nippon     GCATCAAGCACACTACAAAGTAGCTCACAACGCCTTGCTTAACCACGCCCCCACGGGAAA
R.pumilus                  GCATCAAGCGCACTATAAAGTAGCTCATGACGCCTTGCTTAACCACGCCCCCACGGGAAA
```

```
R.pusillus                 GTATCAAGCACACTATAAAGTAGCTCATGACGCCTTGCTTAACCACACCCCCACGGGAAA
R.monoceros                GCATCAAGCACACTATAAAGTAGCTCATGACGCCTTGCTTAACCACACCCCCACGGGAAA
R.macrotis                 GCATCAAGCACACTACAAAGTAGCTCATGACGCCTTGCTTAACCACACCCCCACGGGAAA
R.rex                      GCATCAAGCACACTACAAAGTAGCTCATGACGCCTTGCTTAACCACGCCCCCACGGGAAA
R.affinis                  GCATCAAGCACACTACAAAGTAGCTCATGACGCCTTGCTTAACCACGCCCCCACGGGAAA
RaTG13-sample              GCATCAAGCACACTACAAAGTAGCTCATGACGCCTTGCTTAACCACGCCCCCACGGGAAA
R.sinicus-sinicus          GCATCAAGCACACTACAAAGTAGCTCATGACGCCTTGCTTAACCACACCCCCACGGGAAA
R.thomasi                  GCATCAAGCACACTATAAAGTAGCTCATAACGCCTTGCTTAGCCACACCCCCACGGGAAA
                           * ******* ***** ***********  ************ **** *************

R.yunnanensis              CAGCAGTGATAAAAATTAAGCCATGAACGAAAGTTTGACTAAGTTATACCTACTCCTTAG
R.ferrumequinum-nippon     CAGCAGTGATAAAAATTAAGCCATGAACGAAAGTTTGACTAAGTTATACCTAC-CCCCAG
R.pumilus                  CAGCAGTGATAAAAATTAAGCCATGAACGAAAGTTCGACTAAGTTATACCTACTCCTTAG
R.pusillus                 CAGCAGTGATAAAAATTAAGCCATGAACGAAAGTTCGACTAAGTTATACCTACTCCTTAG
R.monoceros                CAGCAGTGATAAAAATTAAGCCATGAACGAAAGTTCGACTAAGTTATACCTACTCCTTAG
R.macrotis                 CAGCAGTGATAAAAATTAAGCCATGAACGAAAGTTCGACTAAGTTATACCTACTCCCCAG
R.rex                      CAGCAGTGATAAAAATTAAGCCATGAACGAAAGTTCGACTAAGTTATACCTACTCCCCAG
R.affinis                  CAGCAGTGATAAAAATTAAGCCATGAACGAAAGTTCGACTAAGTTATACCTACTCCTTAG
RaTG13-sample              CAGCAGTGATAAAAATTAAGCCATGAACGAAAGTTCGACTAAGTTATACCTACTCCTTAG
R.sinicus-sinicus          CAGCAGTGATAAAAATTAAGCCATGAACGAAAGTTCGACTAAGTTATACCTATTCCTTAG
R.thomasi                  CAGCAGTGATAAAAATTAAGCCATGAACGAAAGTTCGACTAAGTTATACCTACTTCCTAG
                           *********************************** ****************   *  **

R.yunnanensis              GGTTGGTAAATTTCGTGCCAGCCACCGCGGTCACACGATTAACCCAAATCAACAGAAACA
R.ferrumequinum-nippon     GGTTGGTAAATTTCGTGCCAGCCACCGCGGTCACACGATTAACCCAAATTAACAGAAATA
R.pumilus                  GGTTGGTAAATTTCGTGCCAGCCACCGCGGTCACACGATTAACCCAAATCAACAGAAACA
R.pusillus                 GGTTGGTAAATTTCGTGCCAGCCACCGCGGTCACACGATTAACCCAAATCAACAGAAACA
R.monoceros                GGTTGGTAAATTTCGTGCCAGCCACCGCGGTCACACGATTAACCCAAATCAACAGAAATA
R.macrotis                 GGTTGGTAAATTTCGTGCCAGCCACCGCGGTCACACGATTAACCCAAATCAACAGAAACA
R.rex                      GGTTGGTAAATTTCGTGCCAGCCACCGCGGTCACACGATTAACCCAAATCAACAGAAACA
R.affinis                  GGTTGGTAAATTTCGTGCCAGCCACCGCGGTCACACGATTAACCCAAATTAACAGAAACA
RaTG13-sample              GGTTGGTAAATTTCGTGCCAGCCACCGCGGTCACACGATTAACCCAAATTAACAGAAACA
R.sinicus-sinicus          GGTTGGTAAATTTCGTGCCAGCCACCGCGGTCACACGATTAACCCAAATCAACAGAAATA
R.thomasi                  GGTTGGTAAATTTCGTGCCAGCCACCGCGGTCACACGATTAACCCAAATCAACAGAAATA
                           ************************************************* ******** *

R.yunnanensis              CGGCGTAAAGCGTGTTTAAGAATAC-AAGAAAAATAAAGTTAAACTCTAGCTAAGCCGTA
R.ferrumequinum-nippon     CGGCGTAAAGCGTGTTTAAGAGTAC---AAAAAATAAAGTTAAATCCTAACTAAGCCGTA
R.pumilus                  CGGCGTAAAGCGTGTTTAAGAATAA--AAAAAAATAAAGTTAAATTCTAGCTAAGCTGTA
R.pusillus                 CGGCGTAAAGCGTGTTTAAGAATAA-AAAAAAAATAAAGTTAAATTCTAGCTAAGCTGTA
R.monoceros                CGGCGTAAAGCGTGTTTAAGAATAA-AAAAAAAATAAAGTTAAATTCTAGCTAAGCTGTA
R.macrotis                 CGGCGTAAAGCGTGTTTAAGAATAATAAAAAAAATAAAGTTAAATTCTAGCTAAGCTGTA
R.rex                      CGGCGTAAAGCGTGTTTAAGAATAATAAAAAAAATAAGGTTAAATTCTAACTAAGCTGTA
R.affinis                  CGGCGTAAAGCGTGTTTAAGAATAC-AAAAAAAATAAAGTTAAATTCTAGCTAAGCTGTA
RaTG13-sample              CGGCGTAAAGCGTGTTTAAGAATACAAAAAAAAATAAAGTTAAATTCTAGCTAAGCTGTA
R.sinicus-sinicus          CGGCGTAAAGCGTGTTTAAGAGTGC--AAAAAAATAAAGTTAAATTCTAGCTAAGCCGTA
R.thomasi                  CGGCGTAAAGCGTGTTTAAGAACAT--AAAAAAATAAAGTTAAATTCTAGCTAAGCCGTA
                           *********************        ******** ******  *** ****** ***

R.yunnanensis              AAAAGCCATAGCTAAAATAAAAATAGACCACGAAAGTGACTTTATAAGTTCTGAATACAC
R.ferrumequinum-nippon     AAAAGCCCTAGCTAAAATAAAAATAAACTACGAAAGTGACTTTACGAATTCTGAATACAC
R.pumilus                  AAAAGCCATAGCTAAAATAAAAATAGACCACGAAAGTGACTTTACAAGTTCTGAATACAC
R.pusillus                 AAAAGCCATAGCTAAAATAAAAATAGACCACGAAAGTGACTTTACAAGTTCTGAGTACAC
R.monoceros                AAAAGCCATAGCTAAAATAAAAATAGACCACGAAAGTGACTTTACAAGTTCTGAGTACAC
R.macrotis                 AAAAGCCATAGCTAAAATAAAAATAAACCACGAAAGTGACTTTACAAATTCTGAATACAC
R.rex                      AAAAGCCATAGCTAAAATAAAAATAAACTACGAAAGTGACTTTACAAATTCTGAATACAC
R.affinis                  AAAAGCCATAGCTAAAATAAAAATAGACTACGAAAGTGACTTTACAAATTCTGAATACAC
RaTG13-sample              AAAAGCCATAGCTAAAATAAAAATAGACTACGAAAGTGACTTTACAAATTCTGAATACAC
R.sinicus-sinicus          AAAAGCCATAGCTAAAATAAAAATAAACTACGAAAGTGACTTTACGAATTCTGAACACAC
R.thomasi                  AAAAGCCATAGCTAAAATAAAAATAAACTACGAAAGTGACTTTACGAATTCTGAACACAC
                           ******* ***************** ** ***************  * ******  ****

R.yunnanensis              GATAGCTAAGACCCAAACTGGGATTAGATACCCCACTATGCTTAGCCCTAAACCTAAACA
R.ferrumequinum-nippon     GATAGCTAAGACCCAAACTGGGATTAGATACCCCACTATGCTTAGCCCTAAACCTAAACA
R.pumilus                  GATAGCTAAGACCCAAACTGGGATTAGATACCCCACTATGCTTAGCCCTAAACCTAAACA
R.pusillus                 GATAGCTAAGACCCAAACTGGGATTAGATACCCCACTATGCTTAGCCCTAAACCTAAACA
R.monoceros                GATAGCTAAGACCCAAACTGGGATTAGATACCCCACTATGCTTAGCCCTAAACCTAAACA
R.macrotis                 GATAGCTAAGATCCAAACTGGGATTAGATACCCCACTATGCTTAGCCCTAAACCTAAACA
```

```
R.rex                      GATAGCTAAGACCCAAACTGGGATTAGATACCCCACTATGCTTAGCCCTAAACCTAAACA
R.affinis                  GATAGCTAAGACCCAAACTGGGATTAGATACCCCACTATGCTTAGCCCTAAACCTAAACA
RaTG13-sample              GATAGCTAAGACCCAAACTGGGATTAGATACCCCACTATGCTTAGCCCTAAACCTAAACA
R.sinicus-sinicus          GATAGCTAAGACCCAAACTGGGATTAGATACCCCACTATGCTTAGCCCTAAACCTAAACA
R.thomasi                  GATAGCTAAGACCCAAACTGGGATTAGATACCCCACTATGCTTAGCCCTAAACCTAAACA
                           *********** ************************************************

R.yunnanensis              ATCAACACAACAACATTATTCGCCAGAGTACTACCAGCAACAGCTTAAAACTCAAAGGAC
R.ferrumequinum-nippon     ATCAACACAACAACATTGTTCGCCAGAGTACTACTAGCAATAGCTTAAAACTCAAAGGAC
R.pumilus                  ATTAACACAACAATATTATTCGCCAGAGTACTACTAGCAATAGCTTAAAACTCAAAGGAC
R.pusillus                 ATTAGCACAACAATATTATTCGCCAGAGTACTACTAGCAACAGCTTAAAACTCAAAGGAC
R.monoceros                ATTAGCACAACAATATTATTCGCCAGAGTACTACTAGCAACAGCTTAAAACTCAAAGGAC
R.macrotis                 ATCAACACAACAACATTATTCGCCAGAGTACTACTAGCAACAGCTTAAAACTCAAAGGAC
R.rex                      ATCAACACAACAACATTATTCGCCAGAGTACTACTAGCAACAGCTTAAAACTCAAAGGAC
R.affinis                  ATCAACACAACAACATTATTCGCCAGAGTACTACTAGCAACAGCTTAAAACTCAAAGGAC
RaTG13-sample              ATCAACACAACAACATTATTCGCCAGAGTACTACTAGCAACAGCTTAAAACTCAAAGGAC
R.sinicus-sinicus          ATCAACACAACAACATTATTCGCCAGAGTACTACTAGCAACAGCTTAAAACTCAAAGGAC
R.thomasi                  GTCAACACAACAACATTATTCGCCAGAGTACTACTAGCAACAGCTTAAAACTCAAAGGAC
                            * * ******** *** **************** ***** *******************

R.yunnanensis              TTGGCGGTGCTTTATACCCCTCTAGAGGAGCCTGTCCTGTAATCGATAGACCCCGATAAA
R.ferrumequinum-nippon     TTGGCGGTGCTTTATACCCCTCTAGAGGAGCCTGTCCTATAATCGATAAACCCCGATAGA
R.pumilus                  TTGGCGGTGCTTTATACCCCTCTAGAGGAGCCTGTCCTATAATCGATAAACCCCGATAGA
R.pusillus                 TTGGCGGTGCTTTATACCCCTCTAGAGGAGCCTGTCCTATAATCGATAAACCCCGATAGA
R.monoceros                TTGGCGGTGCTTTATACCCCTCTAGAGGAGCCTGTCCTATAATCGATAAACCCCGATAGA
R.macrotis                 TTGGCGGTGCTTTATACCCCTCTAGAGGAGCCTGTCCTATAATCGATAAACCCCGATAGA
R.rex                      TTGGCGGTGCTTTATACCCCTCTAGAGGAGCCTGTCCTATAATCGATAAACCCCGATAAA
R.affinis                  TTGGCGGTGCTTCATACCCCTCTAGAGGAGCCTGTCCTATAATCGATAAACCCCGATAGA
RaTG13-sample              TTGGCGGTGCTTCATACCCCTCTAGAGGAGCCTGTCCTATAATCGATAAACCCCGATAAA
R.sinicus-sinicus          TTGGCGGTGCTTTATACCCCTCTAGAGGAGCCTGTCCTATAATCGATAAACCCCGATAGA
R.thomasi                  TTGGCGGTGCTTTATACCCCTCTAGAGGAGCCTGTCCTATAATCGATAAACCCCGATAGA
                           ************ ************************* ********* ********* *

R.yunnanensis              CCTCACCAGCCCTTGCCAACTCAGCTTATATACCGCCATCCTCAGCAAACCCTAAAAAGG
R.ferrumequinum-nippon     CCTCACCAGCCCTTGCCAATTCAGCCTATATACCGCCATCCCCAGTAAACCCTAAAAAGG
R.pumilus                  CCTCACCAGCTCTTGCCAATTCAGCCTATATACCGCCATCCTCAGCAAACCCTAAAAAGG
R.pusillus                 CCTCACCAGCTCTTGCCAATTCAGCCTATATACCGCCATCCTCAGCAAACCCTAAAAAGG
R.monoceros                CCTCACCAGCTCTTGCCAATTCAGCCTATATACCGCCATCCTCAGCAAACCCTAAAAAGG
R.macrotis                 CCTCACCAGCTCTTGCCAATTCAGCCTATATACCGCCATCCTCAGCAAACCCTAAAAAGG
R.rex                      CCTCACCAGCTCTTGCCAACTCAGCTTATATACCGCCATCCTCAGCAAACCCTAAAAAGG
R.affinis                  CCTCACCAGCTCTTGCCAATTCAGCCTATATACCGCCATCCTCAGCAAACCCTAAAAAGG
RaTG13-sample              CCTCACCAGCTCTTGCCAATTCAGCCTATATACCGCCATCCTCAGCAAACCCTAAAAAGG
R.sinicus-sinicus          CCTCACCAGCTCTTGCCAATTCAGCTTATATACCGCCATCCTCAGCAAACCCTAAAAAGG
R.thomasi                  CCTCACCAGCTCTTGCCAATTCAGCTTATATACCGCCATCCTCAGCAAACCCTAAAAAGG
                           ********** ******** ***** *************** *** **************

R.yunnanensis              AGCCACAGTAAGCACAAACATAAGACATAAAAACGTTAGGTCAAGGTGTAGCCCATGGGC
R.ferrumequinum-nippon     AACTGTAGTAAGCACAAACATAAGACATAAAGACGTTAGGTCAAGGTGTAGCCCATGAGC
R.pumilus                  AACTGTAGTAAGCACAAACATAAGACATAAAAACGTTAGGTCAAGGTGTAGCCTATGAGC
R.pusillus                 AACTGTAGTAAGCACAAACATAAGACATAAAAACGTTAGGTCAAGGTGTAGCCTATGAGC
R.monoceros                AACTGTAGTAAGCACAAACATAAGACATAAAAACGTTAGGTCAAGGTGTAGCCTATGAGC
R.macrotis                 AACTGTAGTAAGCACAAACATAAGACATAAAAACGTTAGGTCAAGGTGTAGCCTATGAGC
R.rex                      AACTGTAGTAAGCACAAACATAAGACATAAAAACGTTAGGTCAAGGTGTAGCCTATGAGC
R.affinis                  AACTGCAGTAAGCACAAACATTAGACATAAAAACGTTAGGTCAAGGTGTAGCCTATGAGC
RaTG13-sample              AACTGCAGTAAGCACAAACATTAGACATAAAAACGTTAGGTCAAGGTGTAGCCTATGAGC
R.sinicus-sinicus          AACTACAGTAAGCACAAACATAAGACATAAAAACGTTAGGTCAAGGTGTAGCCTATGAGC
R.thomasi                  AACTACAGTAAGCACAAACATAAGACATAAAAACGTTAGGTCAAGGTGTAGCCTATGAGC
                           * *   *************** ********* ********************* *** **

R.yunnanensis              TGGGAAGAGATGGGCTACATTTTCTTACCAAAGAACACTTAAAACCACATACGGAAGCTA
R.ferrumequinum-nippon     TGGGAAGAGATGGGCTACATTTTCTTATCAAAGAACACTTAAAATTTCATACGGAAACTC
R.pumilus                  TGGGAAGAGATGGGCTACATTTTCTTCTCAAAGAACACTTAAAATCATATACGAAAGCTC
R.pusillus                 TGGGAAGAGATGGGCTACATTTTCTTCTCAAAGAACACTTAAAATCATATACGAAAGCTC
R.monoceros                TGGGAAGAGATGGGCTACATTTTCTTCTCAAAGAACACTTAAAATCATATACGAAAGCTC
R.macrotis                 TGGGAAGAGATGGGCTACATTTTCTTCTCAAAGAACACTTAAAGTCACATACGGAAGCTC
R.rex                      TGGGAAGAGATGGGCTACATTTTCTTCTTAAAGAACACTTAAAATTACATACGGAAGCTC
R.affinis                  TGGGAAGAGATGGGCTACATTTTCTTCTCAAAGAACATTTAAAACTACATACGGAAGTTC
RaTG13-sample              TGGGAAGAGATGGGCTACATTTTCTTCTCAAAGAACATTTAAAACTACATACGGAAGTTC
```

```
R.sinicus-sinicus           TGGGAAGAGATGGGCTACATTTTCTTCTTAAAGAACACTTAAAACTTCATACGGAAGCTC
R.thomasi                   TGGGAAGAGATGGGCTACATTTTCTTCTTAAAGAACATTTAAAACTTCATACGGAAGCTC
                            **************************   ******** *****     ***** **  *

R.yunnanensis               CCATGAAACAAAGAGCGGAAGGTGGATTTAGTAGTAAATCAAGAATAAAGAGCTTGATTG
R.ferrumequinum-nippon      CCATGAAACGGGGAGCAGAAGGTGGATTTAGTAGTAAACCAAGAACAAAGAGCTTGGTTG
R.pumilus                   CCATGCAACAGGGAGCGGAAGGTGGATTTAGTAGTAAACCAAGAACAAAGAGCTTGATTG
R.pusillus                  CCATGCAACAGAGAGCGGAAGGTGGATTTAGTAGTAAACCAAGAACAAAGAGCTTGATTG
R.monoceros                 CCATGCAACAGAGAGCGGAAGGTGGATTTAGTAGTAAACCAAGAACAAAGAGCTTGATTG
R.macrotis                  CCATGCAACAGGGGGCGGAAGGTGGATTTAGTAGTAAACCAAGAACAAAGAGCTTGATTG
R.rex                       CCATGCAATAGGGGGCAGAAGGTGGATTTAGTAGTAAACCAAGAACAAAGAGCTTGATTG
R.affinis                   TCATGAAATAGAGAGCGGAAGGTGGATTTAGTAGTAAATCAAGAACAAAGAGCTTGGTTG
RaTG13-sample               TCATGAAATAGAGAACGGAAGGTGGATTTAGTAGTAAATCAAGAACAAAGAGCTTGGTTG
R.sinicus-sinicus           CAATGAAATAAAGAGCGGAAGGTGGATTTAGTAGTAAATCAAGAACAAAGAGCTTGATTG
R.thomasi                   CCGTGAAACAAGGAGCAGAAGGTGGATTTAGTAGTAAATCAAGAACAAAGAGCTTGATTG
                               ** **    *  * ********************* ****** ********** ***

R.yunnanensis               AATCAGGCCATGAAGCACGCACACACCGCCCGTCACCCTCCTCAAATACAAAGGAAGTGC
R.ferrumequinum-nippon      AATTAGGCCATGAAGCACGCACACACCGCCCGTCACCCTCCTCAAATATAAAGGTACCAC
R.pumilus                   AATTAGGCCATGAAGCACGCACACACCGCCCGTCACCCTCCTCAAATATAGAGGTAATAC
R.pusillus                  AATTAGGCCATGAAGCACGCACACACCGCCCGTCACCCTCCTCAAATATAGAGGTAGCAC
R.monoceros                 AATTAGGCCATGAAGCACGCACACACCGTCCGTCACCCTTCTCAAATATAGAGGTAGCAC
R.macrotis                  AATTAGGCCATGAAGCACGCACACACCGCCCGTCACCCTCCTCAAATACGGAGGTAATAC
R.rex                       AATTAGGCCATGAAGCACGCACACACCGCCCGTCACCCTCCTCAAATATAAAGGTAGTAC
R.affinis                   AATTAGGCCATGAAGCACGCACACACCGCCCGTCACCCTCCTCAAATATGAAGGTAATAC
RaTG13-sample               AATTAGGCCATGAAGCACGCACACACCGCCCGTCACCCTCCTCAAATATGAAGGTAATGC
R.sinicus-sinicus           AATAAGGCCATGAAGCACGCACACACCGCCCGTCACCCTCCTCAAATATAGAGGTAATAC
R.thomasi                   AATTAGGCCATGAAGCACGCACACACCGCCCGTCACCCTCCTCAAATATAGAGGTAGCAC
                            *** ************************ ********** ********   *** *   *

R.yunnanensis               CCAAACCTATTACCATACACCCACAGTATGAGAGGAGATAAGTCGTAACAAGGTAAG-CG
R.ferrumequinum-nippon      CCAAACCTATTAACACGTACCCACAACATGAGAGGAGATAAGTCGTAACAAGGTAAG-CG
R.pumilus                   CCAAACCTATTATCACGTACCCATAGTATGAGAGGAGATAAGTCGTAACAAGGTAAGCCG
R.pusillus                  CCAAACCTATTACCACGTACCCATAGTATGAGAGGAGATAAGTCGTAACAAGGTAAG-CG
R.monoceros                 CCAAACCTATTACCACGTACCCATAGTATGAGAGGAGATAAGTCGTAACAAGGTAAG-CG
R.macrotis                  CCAAACCTATTAACACGTGCCCGTAGTATGAGAGGAGATAAGTCGTAACAAGGTAAG-CG
R.rex                       CCAAACCTATTAACACGTACCCGTAGTATGAGAGGAGATAAGTCGTAACAAGGTAAG-CG
R.affinis                   CCAAACCTATTACCACACACCCACAATATGAGAGGAGATAAGTCGTAACAAGGTAAG-CG
RaTG13-sample               CCAAACCTATTACCACACACCCACAATATGAGAGGAGATAAGTCGTAACAAGGTAAG-CG
R.sinicus-sinicus           CCAAACCTATTACCACGTACCCGTAATATGAGAGGAGATAAGTCGTAACAAGGTAAG-CG
R.thomasi                   CTAAACCTATTATCACGTACCCGTAATATGAGAGGAGATAAGTCGTAACAAGGTAAG-CG
                            * ********** **    ***  *  ****************************** **

R.yunnanensis               TACTGGAAAGTGCGCTTGGATACAC-----------------------------------
R.ferrumequinum-nippon      TACTGGAAAGTGCGCTTGGATATAC-----------------------------------
R.pumilus                   TACTGGAAAGTGCGCTTGGATATAT-----------------------------------
R.pusillus                  TACTGGAAAGTGCGCTTGGATATAT-----------------------------------
R.monoceros                 TACTGGAAAGTGCGCTTGGATATAT-----------------------------------
R.macrotis                  TACTGGAAAGTGCGCTTGGATATAT-----------------------------------
R.rex                       TACTGGAAAGTGCGCTTGGATATAC-----------------------------------
R.affinis                   TACTGGAAAGTGCGCTTGGATACAT-----------------------------------
RaTG13-sample               TACTGGAAAGTGCGCTTGGATACATCAAAGTGTAGCTTAAGCCAAAAGCACCTGGCTTAC
R.sinicus-sinicus           TACTGGAAAGTGCGCTTGGATACAT-----------------------------------
R.thomasi                   TACTGGAAAGTGCGCTTGGATACAT-----------------------------------
                            ********************** *

R.yunnanensis               ------------------------------------------------------------
R.ferrumequinum-nippon      ------------------------------------------------------------
R.pumilus                   ------------------------------------------------------------
R.pusillus                  ------------------------------------------------------------
R.monoceros                 ------------------------------------------------------------
R.macrotis                  ------------------------------------------------------------
R.rex                       ------------------------------------------------------------
R.affinis                   ------------------------------------------------------------
RaTG13-sample               ACCCAGGAGACTTCACGTACAATGAACGCTTTGAACAAGTACTAGCCCAACCACAACCCA
R.sinicus-sinicus           ------------------------------------------------------------
R.thomasi                   ------------------------------------------------------------
```

**Supplementary Table 1** Accession numbers of mitochondrial SSU rRNA sequences used in the phylogenetic analysis

| Species | NCBI accession number |
|---|---|
| *R.affinis* | MT845219.1 |
| *R.sinicus sinicus* | KP257597.1 |
| *R.pusillus* | NC_046021.1 |
| *R.yunnanensis* | NC_036419.1 |
| *R.ferrumequinum* | KT779432.1 |
| *R.rex* | KT599913.1 |
| *R.macrotis* | KP162343.1 |
| *R.pumilus* | AB061526.1 |
| *R.thomasi* | KY124333.1 |
| *R.monoceros* | AF406806.1 |